\documentclass[review]{elsarticle}

\usepackage{hyperref}
\usepackage{wasysym}
\usepackage{textcomp}
\usepackage{nicefrac}
\usepackage{upgreek}
\usepackage{lineno}

\journal{Nuclear Instruments and Methods A}

\newcommand{\cmsq}{$\rm{cm}^2$}
\newcommand{\mmsq}{$\rm{mm}^2$}
\newcommand{\mum}{$\upmu \rm{m}$}
\newcommand{\mumsq}{${\upmu \rm{m}}^2$}

\begin{document}

\begin{frontmatter}

\title{Advanced Methods for the Optical Quality Assurance of Silicon Sensors}

\author[tue]{E. Lavrik\corref{cor1}}
\ead{Evgeny.Lavrik@uni-tuebingen.de}
\author[tue]{I. Panasenko}
\author[tue,gsi]{H.R. Schmidt}

\address[tue]{University of T\"ubingen, Auf der Morgenstelle 14, 72076 T\"ubingen, Germany}
\address[gsi]{GSI Helmholtzzentrum f\"ur Schwerionenforschung GmbH, Planckstrasse 1, 64291 Darmstadt, Germany}
\cortext[cor1]{Corresponding author}

\begin{abstract}

We describe a setup for optical quality assurance of silicon microstrip sensors. Pattern recognition algorithms were developed to analyze microscopic scans of the sensors for defects. It is shown that the software has a recognition and classification rate of $>$~90\% for defects like scratches, shorts, broken metal lines etc. We have demonstrated that advanced image processing based on neural network techniques is able to further improve the recognition and defect classification rate.
\end{abstract}


\end{frontmatter}

\section{Introduction}

Rigorous Quality Assurance (QA) of detector components is the key for a successful commissioning of experiments. For large experiments with thousands of identical detector components, the development of automated procedures to detect defects of these components is of critical importance. We have, in the framework of the \textgravedbl Compressed Baryonic Matter\textgravedbl\ Experiment (CBM)~\cite{CBMBook2011}, developed advanced methods to detect and classify defects of silicon microstrip sensors employing optical QA.

CBM is one of the four large experiments under construction at the future international accelerator center Facility for Anti-Proton and Ion Research (FAIR)~\cite{FAIRBTR, FAIRGP} in Darmstadt. The key detector of the CBM experiment is a complex, multi-layer Silicon Tracking System (STS)~\cite{STSTDR}. The performance requirements are (a) pile-up free track measurement at the anticipated high collision rates of up to 10~MHz, (b) a momentum resolution of better than 2\% in a 1~Tm dipole magnetic field, and (c) capabilities for the identification of particle decays with displaced vertices, e.g., those with strangeness content.

To meet these requirements, it is necessary to restrict the material budget in the acceptance to an absolute minimum. The basic building block of the STS is a module, which consists of a double-sided silicon microstrip sensor and a multi-layer microcable which connects the sensor to the front-end electronics (FEE). The sensor sizes used in the setup are $6.2 \times 2.2$~\cmsq, $6.2 \times 4.2$~\cmsq, $6.2 \times 6.2$~\cmsq and $6.2 \times 12.4$~\cmsq. 
The wafer material is of n-type. The sensors are segmented into 1024 strips per side at a strip pitch of 58~\mum. The strips are read out through integrated AC coupling. The p-strips are arranged under a stereo angle of $7.5^{\circ}$ with respect to the n-strips. The short corner strips are interconnected using a second metal layer in order to enable full readout of the p-side from one sensor edge only, like with the simpler topology of the n-side. To degrade the field towards edge, the sensor has multiple and single guards ring on the n-and p-side, respectively.

\begin{figure}[!htb]
	\hspace*{\fill}
	\includegraphics[width=.48\textwidth]{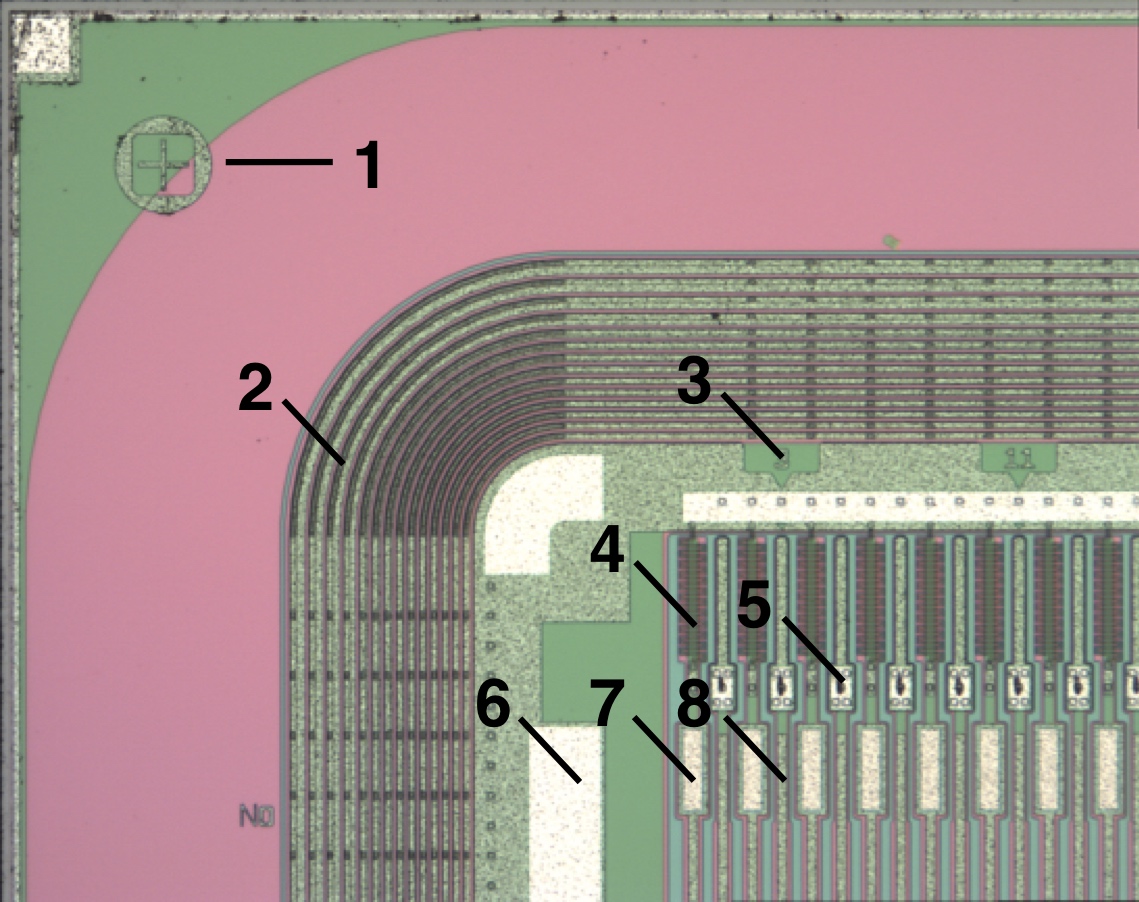}
	\hfill
	\includegraphics[width=.48\textwidth]{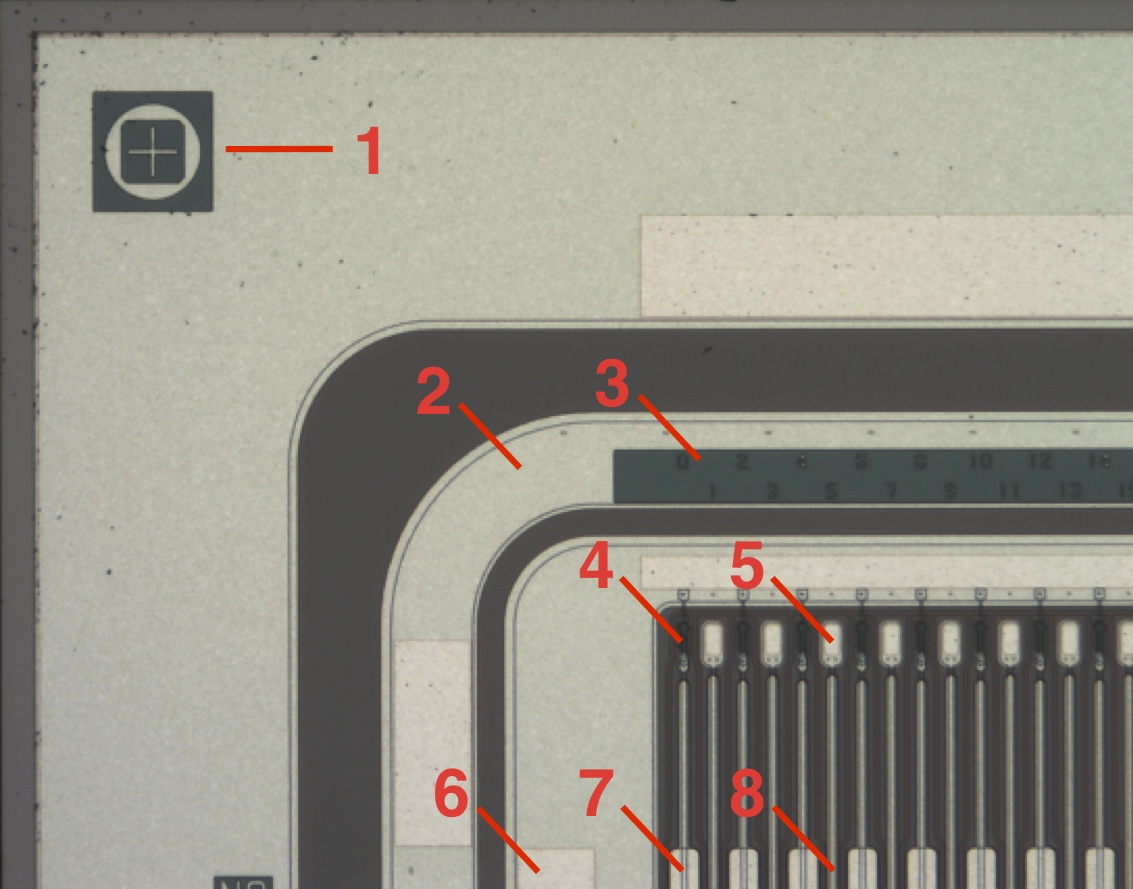}
	\hspace*{\fill}
	\caption{Layout of the silicon sensors from CiS (left panel) and Hamamatsu (right panel) in the corner region, n side. Legend: 1) Alignment mark 2) Guard ring 3) Strip numbering panel 4) Polysilicon bias resistor 5) DC pad 6) Bias ring pad 7) AC pad 8) Aluminum read-out strip.}
	\label{fig:sensor-layout}
\end{figure}

The microcables, which connect the sensors to the amplifier ASICs, have a length of up to 50 cm, which is needed to mount the FEE outside of the detector's acceptance. Altogether, about 900 sensors are arranged to form 8 tracking stations.

To ensure the proper functioning of a module, it is mandatory to inspect the sensors both optically~\cite{Lavrik2018Elba} and electrically~\cite{Panasenko2016} \textit{before} bonding the sensors to the microcables. While some electrical tests are part of the vendor's quality assurance to meet the agreed specifications, in-depth optical QA is done exclusively at the client side. The optical QA checks for any damage caused by handling of the sensors and/or for imperfections of the structuring of the sensors. In this contribution we report on a setup to scan sensors microscopically and on the development of advanced methods to detect and analyze defects optically in an automated fashion with minimal human intervention. Including spares, more than 1050 sensors have to be analyzed optically for any defects.

In Sections \ref{Setup} - \ref{Procedures} we describe the setup, calibration and operational prerequisites, respectively, for the optical inspection of the silicon sensors. The subsequent two Sections, \ref{Image_Analysis} and \ref{Advanced_Image_Analysis}, detail the procedures, methods and results of a "conventional" image analysis based on pattern recognition software, and an advanced analysis based on a convolutional deep neural network, respectively. The two methods are compared and a figure of merit for the improvement for the latter on is given.

\section{Mechanical and Optical Setup}
\label{Setup}

\begin{figure}[!htb]
	\centering
	\includegraphics[width=.90\columnwidth]{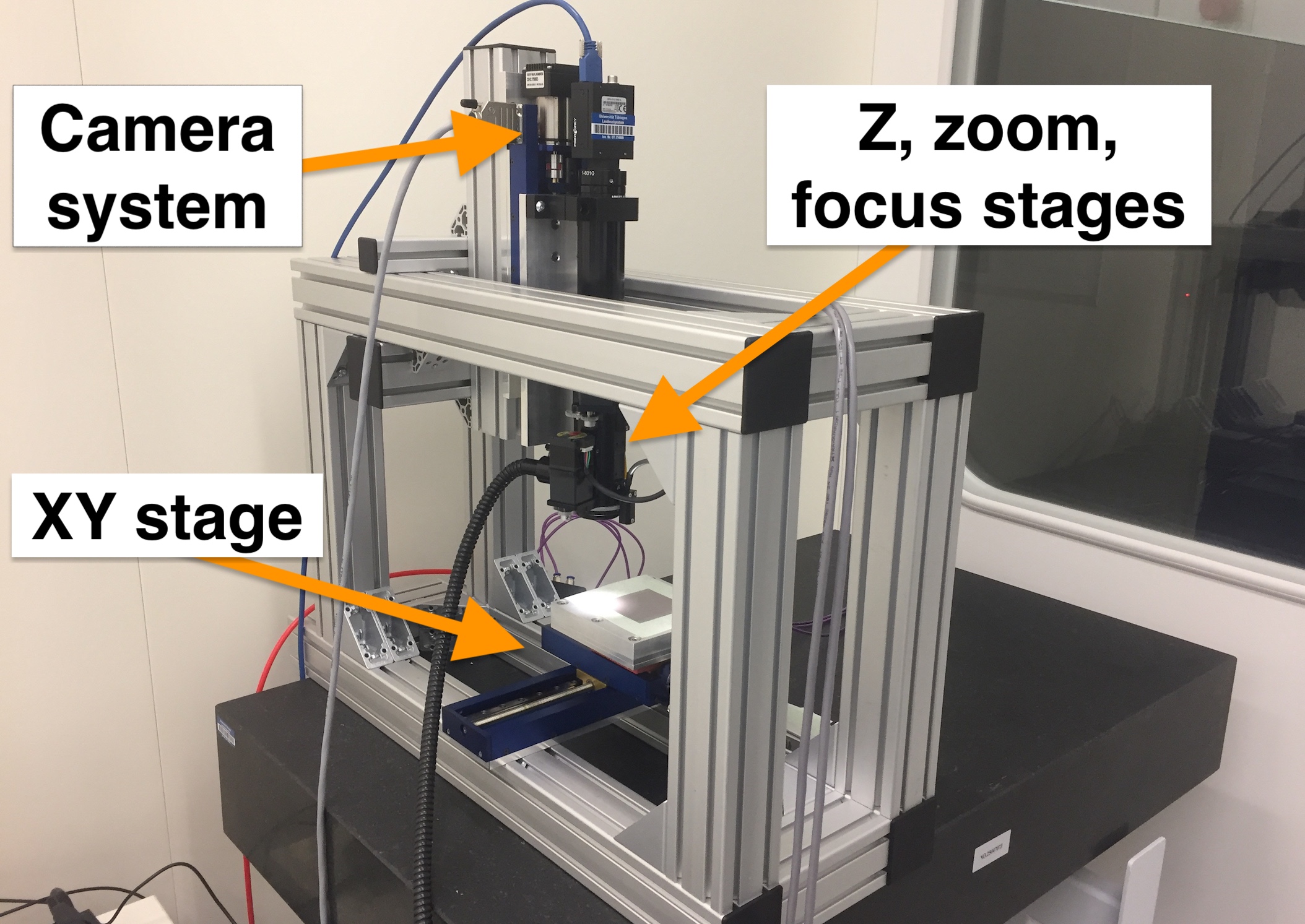}
	\caption{Setup for optical QA.}
	\label{fig:optical-setup-photo}
\end{figure}

The basic components of the optical inspection setup (cf.~Fig.~\ref{fig:optical-setup-photo}) are the camera system mounted on a stable support frame on a massive granite table and an XY-linear stage to move the object under inspection with respect to the camera. A motorized Z-linear stage, the focus and the zoom stages give the setup the flexibility to inspect the object under different focal conditions. The XY-linear table based on Faulhaber/Movtec\textsuperscript{\textregistered} SMC-300 servo motors provides a movement range of 200~mm and 70~mm in X and Y directions, respectively, with a precision of about 0.3~\mum. This allows to mount and inspect sensors up to 124~mm long.

A custom-made 3-zone vacuum chuck allows to mechanically adjust the suction area and underpressure. It is attached to the XY stage to hold the different sensor variants in place during the scanning. A Becker\textsuperscript{\textregistered} 150~mbar vacuum pump supplies the vacuum to the chuck. A vacuum pump bypass channel is controlled by an electromagnetic relay, allowing to release the sensors from the vacuum suction.

The Z-motor stage allows the vertical movement of the optical assembly within a range of 70~mm with a precision of about 0.3~\mum. The optical assembly consists of a motorized 12$\times$ zoom (0.58$\times$ - 7$\times$) and a motorized 3~mm fine focus tube from Navitar\textsuperscript{\textregistered}. The optical tube allows the direct coaxial inline light illumination by a $\diameter$15~mm flexible light guide connected to a 590~lm, 5700~K model Roma LED3 direct light source from Starlight\textsuperscript{\textregistered} company. Use of a ring light source was investigated and phased out due to mirror-like surface of the sensors and not satisfactory image quality. A 5 megapixel microscope camera from Motic is attached to the optical tube with a C-ring connection. The camera has a sensor size of 1/2.5” with a 4:3 aspect ratio and characteristic pixel size of $2.2\times2.2$~\mumsq.

The setup is installed in a clean room with temperature and a humidity control. The required minimum clean room class is ISO 4 (ISO 14644-1 Cleanroom Standards) to avoid excessive distortion of the pattern recognition by dust particles onthe sensor surface.
The setup is controlled with a PC running LabVIEW 2013 software~\cite{Lab01}.

\section{Calibration}
\label{Calibration}
\subsection{Linear Motor Stages}

\begin{figure}[!htb]
	\centering
	\includegraphics[width=.70\columnwidth]{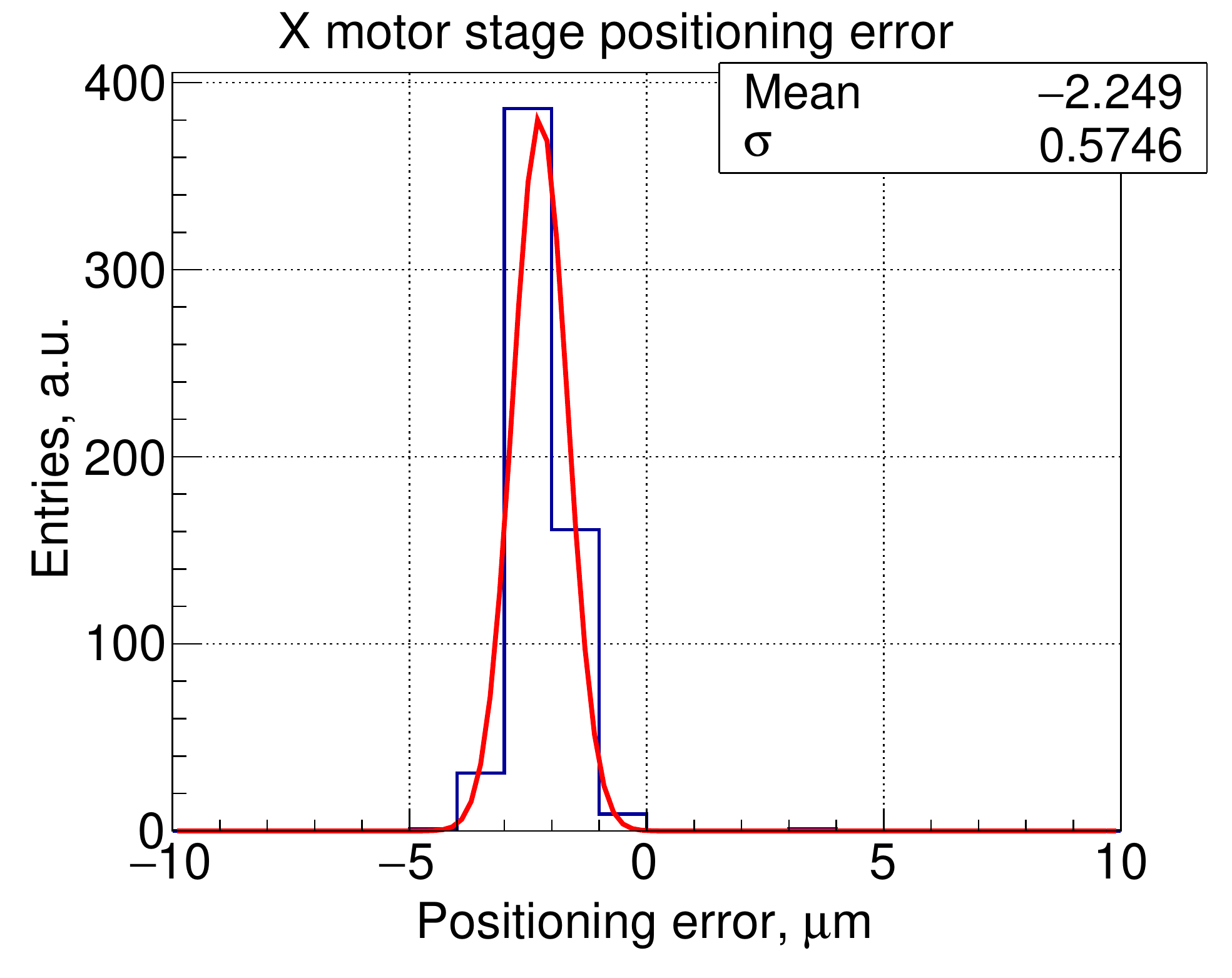}
	\caption{Estimate of the motor stage positioning precision by measuring repeatedly the residuum of actual and targeted position.}
	\label{fig:stages-repeatability}
\end{figure}

The linear motors use closed-loop feedback correction. In order to estimate the uncalibrated motor positioning precision and repeatability, a set of repetitive positioning measurements were carried out. An example of the deviation of the actual position from the target position value is shown in Figure~\ref{fig:stages-repeatability} for the X direction. Here, the actual position was measured immediately upon arrival. This results in a biased distribution, because motor driver reports end of movement before the closed-loop corrections are carried out. The positioning error is further influenced by the jitter of about 0.5 - 1~\mum for all three axes.

\subsection{Optical System}

The optical tube from Navitar with motorized zoom and focus uses conventional spherical lenses. The images taken are thus subject to various geometrical and exposure distortions, e.g., barrel distortions or vignetting. The use of a telecentric objective would mitigate the lens distortion problem, however, the use of auto-focusing (custom implementation with optimizations described in~\cite{Lavrik2017}, based on Fast Fourier Transform) for the high precision contactless height measurements would be impossible.

For a correct and automated recognition and classification of defects, the correction of the distortions is mandatory. We discuss in following the corrections of these distortions in software.

Geometrical barrel distortions are corrected with the help of a dot calibration array. The optimal array dot size and spacing for the present optical system is estimated based on the field-of-view calculations~\cite{Piper1901,Taylor1892}. The individual dots should be visible and detectable at lowest magnification and as many as possible should be visible at highest magnification. Accordingly, a 50~\mum\ dot size and a 100~\mum\ dot spacing has been chosen.

\begin{figure}[!htb]
	\centering
	\includegraphics[width=.7\textwidth]{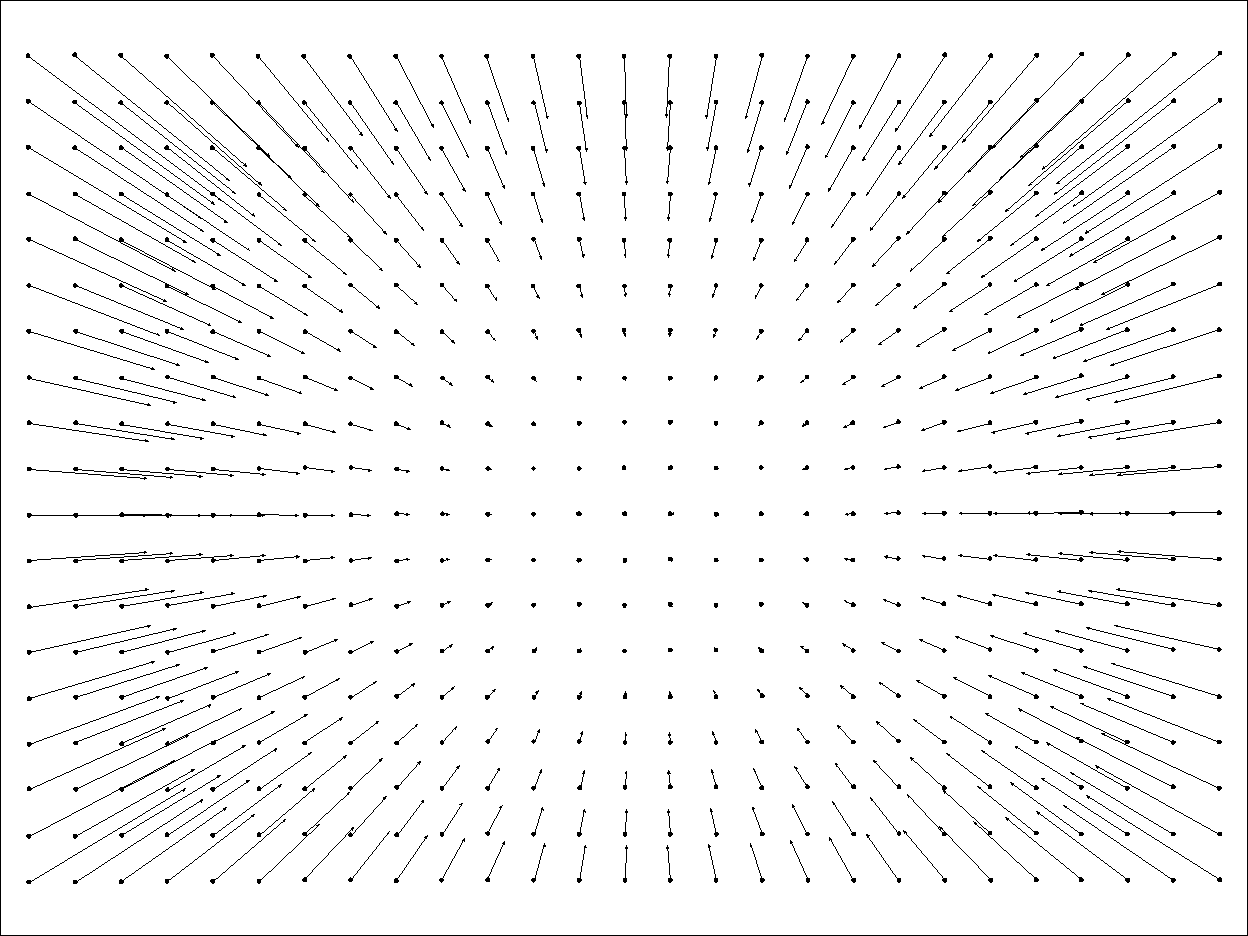}
	\caption{Correction for the barrel distortions of the image. The errors are scaled by a factor of 450 for demonstration purposes. The dimensions of the field are 2.4 by 2.0~\mmsq.}
	\label{fig:axis-correction-error-vector-field}
\end{figure}

The barrel distortion correction is done using the NI Vision software package~\cite{Vis01}. The dots of the calibration plate are recognized by the detection algorithms, their centers and spacings are taken into account. Then, using a polynomial distortion model based on the Brown-Conrady model~\cite{Conrady1919,Brown1966}, the corrected positions are calculated. The residuals from this normalization are represented as the vectors pointing into the direction of the corrected position as shown in Fig.~\ref{fig:axis-correction-error-vector-field}. The vector sizes are scaled by a factor of 450. The maximal value of the correction is  2.1~\mum.

Several approaches exist to correct for vignetting (e.g.,~\cite{Lin2009} or~\cite{Catrysse2000}). In this work we use the \textit{Flat Field Correction} tools from the NI Vision package to account for these effects. It supports several ways of correcting the lighting distortions. The use of mathematical bright field modeling is advised if there is no possibility to acquire an image of the background only.

In order to further improve the flat field correction, the dark field (or dark frame) pattern is measured. It allows to account for the camera gain, pixel pedestals and temporal noise due to the dark currents (covered in~\cite{Porter2008} for CMOS and in~\cite{Widenhorn2002} for CCD sensors). This is done by tightly covering the camera sensor and take digital images in the absence of light. The images taken with the digital camera used in this work show that, in the absence of light, the color channel brightness values do not correspond to a value of 0 (for the black color) but rather to a value range between 12 and 15, which indicates noise in a pixel around the pedestal levels. The computed dark field pattern further improves the flat field correction.

In addition, the optical system's parcentricity, parfocality and magnification are calibrated mechanically using procedures described in~\cite{Lavrik2017}.

The described correction algorithms provide the foundation to configure the optical system to achieve the optimal conditions for acquisition of the best quality images. They have to be carried out once for the chosen inspection conditions - e.g. fixed magnification, illumination and exposure time, etc. - and applied to the grabbed images on the post-processing stage.

\section{Optical Inspection Procedures and Methods}
\label{Procedures}
\subsection {Sensor Alignment and Scanning}

During the placement on the vacuum chuck, the sensor is manually aligned. However, the absence of the rotational motor stage does not allow to align it perfectly. Thus the misalignment of the sensor axes with respect to the motor stage axes must be determined. The key elements for the alignment calibration are several fiducial alignment marks in the corners of the sensor.

The alignment marks are detected with pattern matching algorithms.
Having all the alignment mark positions in the motor coordinate space and knowing the topology of the sensor from the CAD files, it is possible to solve the equation~(\ref{eq:calibration}) to convert between coordinate spaces:

\begin{equation}
\label{eq:calibration}
\vec{x}_m = \mathbf{C} \cdot \mathbf{R} \cdot \left( \vec{t} + \vec{x}_s \right),
\end{equation}

where $\vec{t}$ is the translation vector between the origins of the coordinate spaces, C is the conversion matrix which converts motor steps to micrometers; R is the rotation matrix which extracts the rotation angle between the sensor coordinate space and the motor coordinate space. The 2-dimensional (the Z-direction and roll and pitch angles are not taken into account) expansion of the matrix equation~(\ref{eq:calibration}) is fitted by a ROOT~\cite{Roo01} based macro.
The positioning precision after the calibration is estimated to be $\pm$ 2.4~\mum\ for the X- and $\pm$ 3.1~\mum\ for the Y-motor stage, respectively, over the full movement range. These positioning errors are not critical, since the regions of interest have an overlap of 100~\mum.

Having the sensor misalignment information extracted, the scan along the sensor can be performed. The sensor is divided into a grid of rectangles which are slightly less wide than the camera's field of view at a given magnification. The scan is then performed in a snake-like movement pattern and the images are recorded at each grid point and stored for offline processing. This procedure is subject of optimization: on one hand, the inspection time of a single sensor side should be minimized, on the other hand the magnification of the system should be as high as possible to be able to observe the fine structures of the sensor and possible defects in details. The total system magnification of 1$\times$ was chosen for the sensor scan. In this case a single pixel on the camera sensor corresponds to 1~\mum\ in on the silicon sensor, which is of the same order as the positioning precision.
The corresponding duration for a single scan at the given magnification is about 20 minutes for a $6.2 \times 6.2$~\cmsq sensor, which is the most common sensor type. The scan duration was driven here primarily by the low speed of the camera (about one frame per second). With a high-speed camera the motor movement speed would eventually become the limiting factor.

\section{Image Analysis}
\label{Image_Analysis}

The sensor images were analyzed employing Machine Vision Algorithms (MVA) provided by National Instruments. As we will demonstrate and explain below, these algorithms are powerful tools to detect sensor defects. However, in certain cases, the classification of the defects, e.g., their nature (dust, scratch) and severity, is not fully satisfying. We have further improved the analysis quality by applying neural network techniques (refer to section~\ref{Advanced_Image_Analysis}).

The MVAs are combined together into specialized defect finders which are applied to the regions of interest in parallel. The defect classification is based on the scoring system - the algorithm with a higher detection score wins the classification challenge.

The MVAs include geometrical and color transformations, filtering, pattern and texture matching, as well as advanced morphology. Having enhanced certain features in the microscopic image of a sensor by applying a combination of the above algorithms, the~\textit{Particle Analysis tool} from the NI Vision package can be used to determine, e.g., the size and shape parameters of the defect found. In the following section we will describe the defect detection techniques by means of two examples: dust grains and scratches.

\subsection{Foreign Object and Scratch Detection}

During sensor logistics and handling, the surface of the sensor can accumulate various foreign objects. Mostly, these are dust grains. Though the sensors are handled under clean room conditions and dust grains are removed with a nitrogen gun or a soft brush before inspection, not all of them might be removed. To allow the detection of potentially more severe defects, a reliable identification of dust is mandatory.

\begin{figure}[!htb]
	\hspace*{\fill}
	\includegraphics[width=.45\textwidth]{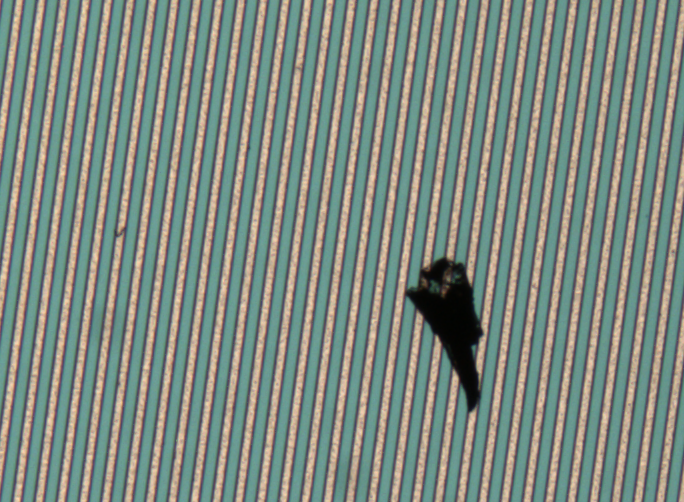}
	\hfill
	\includegraphics[width=.45\textwidth]{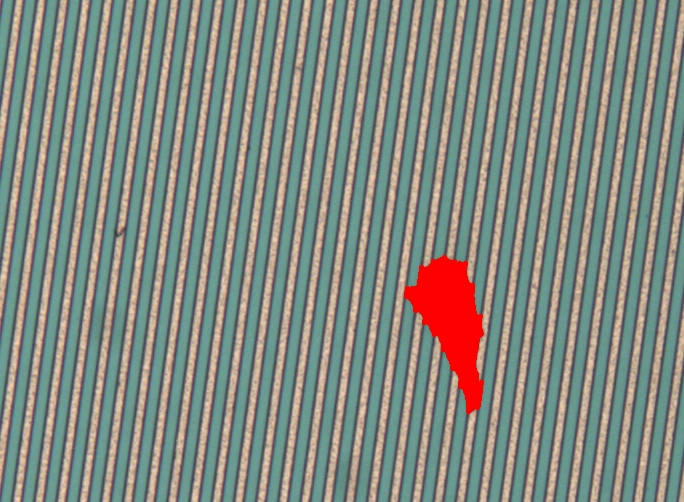}
	\hspace*{\fill}
	\caption{Detection of a dust particle covering 4 strips of a prototype sensor. The left panel shows the source image, the right panel shows the identified defect.}
	\label{fig:defects-dust}
\end{figure}

Figure~\ref{fig:defects-dust} shows an example of dust detection. The left panel presents the source image and the right panel shows the detected defect overlaid onto the source image. Here the luminance plane was extracted and the image was thresholded using a metric auto-threshold. Then, small background objects were removed with the morphology tools. The particle size was measured by the \textit{Particle Analysis tool} and estimated to be 6330~\mumsq. The size of the particles allows to quantify the defect severity.

Scratches are dangerous defects. They can lead to a broken aluminum readout strip. If the scratch is deep enough the implant strip can be affected as well. Both can potentially lead to the breakdown of the strip leading to the saturated readout signal. Moreover the metal and implant strips can be connected by a scratch, which in turn would mimic a pinhole. Another effect is that the metal from a torn strip can electrically shorten the neighboring strips. This results in deterioration of the reconstructed hit position resolution due to the charge redistribution between the readout channels.

\begin{figure}[!htb]
	\hspace*{\fill}
	\includegraphics[width=.45\textwidth]{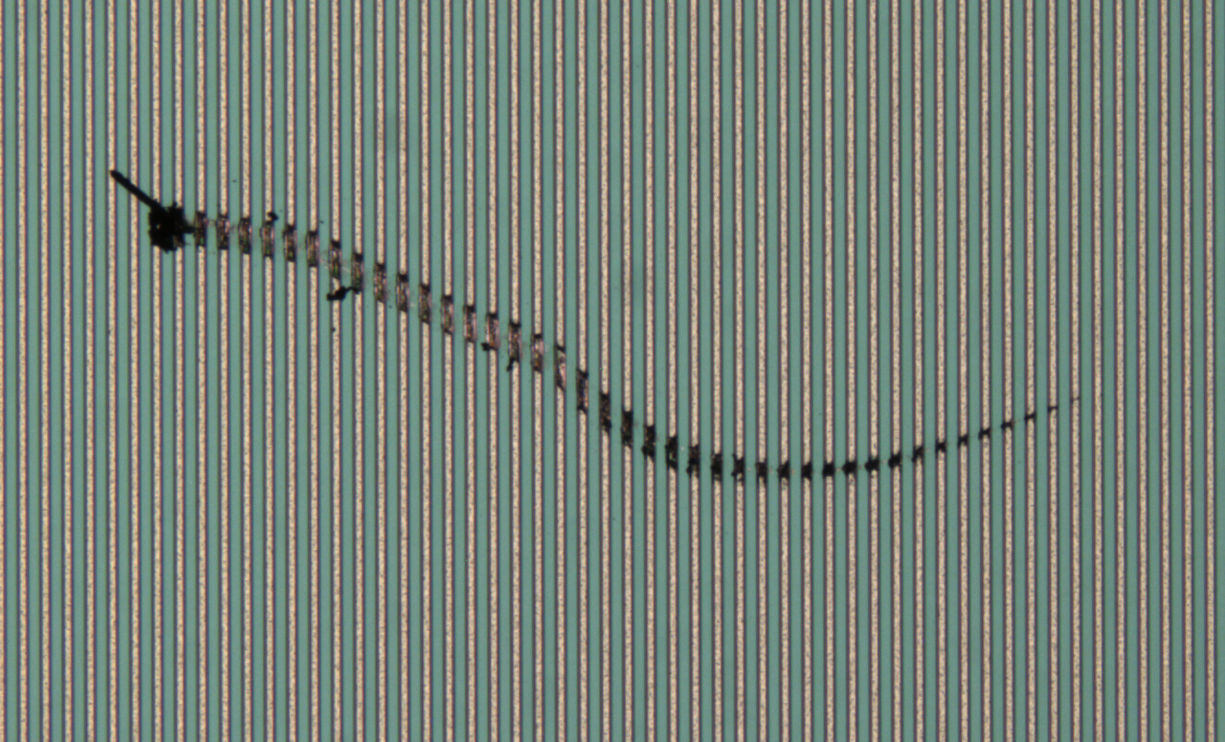}
	\hfill
	\includegraphics[width=.45\textwidth]{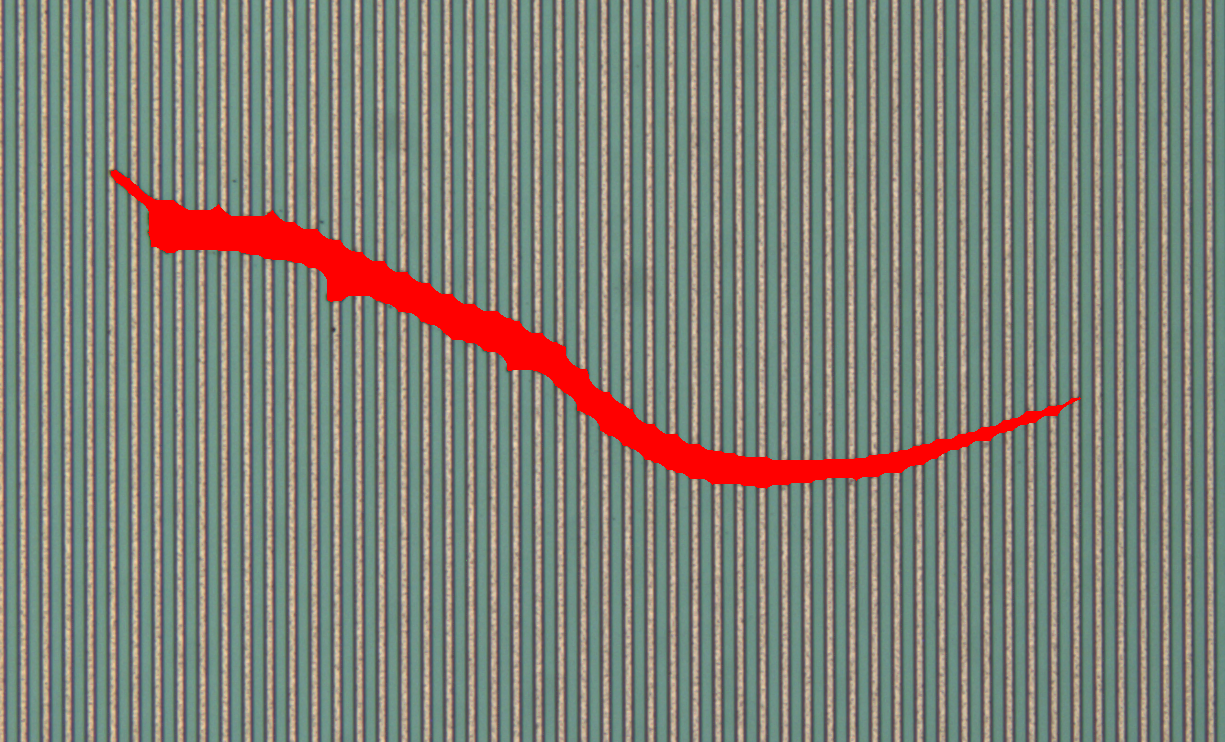}
	\hspace*{\fill}
	\caption{Detection example of a scratch traversing many strips on an n side of a prototype sensor. The left panel shows the source image. The right panel shows the scratch defect detected.}
	\label{fig:defects-scratch}
\end{figure}

Figure~\ref{fig:defects-scratch} shows an example of a scratch detection. The left panel shows the source image and the right panel shows the detected scratch overlaid onto the original image. In this example, the luminance plane of the image was extracted and thresholded manually to extract only the dark objects with brightness values lower than 70 out of 255. The thresholded binary image was then cleaned from small background objects and the interstrip gaps were closed. The resulting single object was analyzed, its total area was measured to be 30130~\mumsq. The analysis tool also indicates a big elongation factor of around 9, which make this defect clearly distinguishable from a typically smaller and more round dust grains. Other metrics of this tool include center of mass, compactness, circularity. perimeter to area ratio, etc. This illustrates how the \textit{Particle Analysis tool} can be used for an offline defect analysis and classification.

\subsection{Strip and Implant Defects}

\begin{figure}[!htb]
	\hspace*{\fill}
	\includegraphics[width=.49\textwidth]{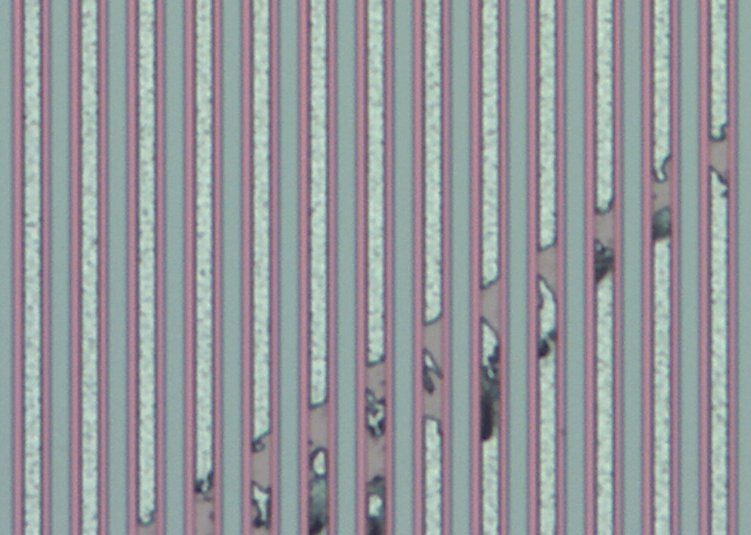}
	\hfill
	\includegraphics[width=.49\textwidth]{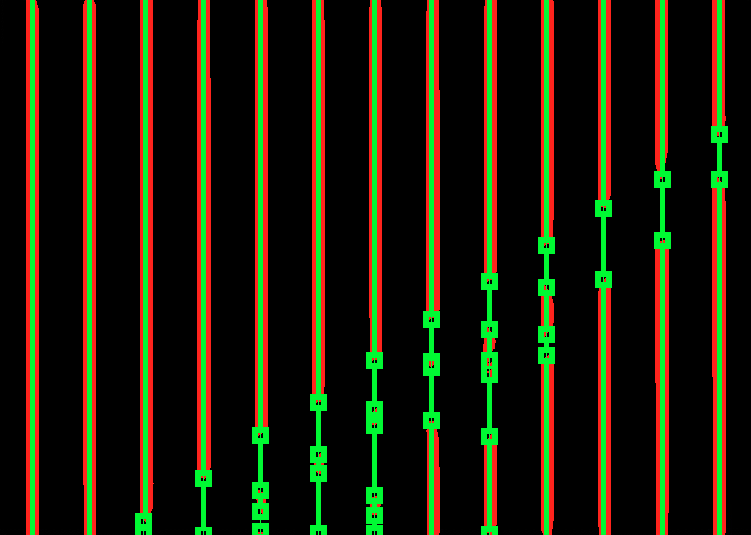}
	\hspace*{\fill}
	\caption{Detection example of a metal break region on a prototype sensor. The left panel shows the source image, the right panel shows the processed image with edges detected, edge numbering is suppressed.}
	\label{fig:defects-metal-open}
\end{figure}

Strip and implant defects are defects on the strip level, i.e., broken aluminum or implant strips.
Figure~\ref{fig:defects-metal-open} shows an example of the detection of defective strips which, like scratches, result in faulty readout channels. This defect, however, occurred during the manufacturing process.
Here, the luminance plane was extracted and the image was then thresholded with a background correction criterium, where the interstrip gap was selected as a background element. The resulting binary image was additionally cleaned up from small objects and the strips were filled with a convex hull morphology algorithm. Then a set of edge detection lines were applied to the strips. For the first two non-defective strips, there are zero edges detected as expected. The defective strips show at least one edge detected, which indicates the defect.

\begin{figure}[!htb]
	\hspace*{\fill}
	\includegraphics[width=.30\textwidth]{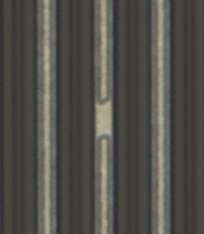}
	\hfill
	\includegraphics[width=.30\textwidth]{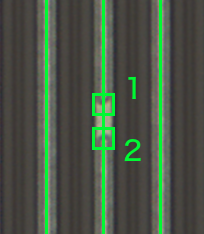}
	\hspace*{\fill}

	\caption{Detection example of an implant strip break. The left panel shows the source image, the right panel shows the detected defect.}
	\label{fig:defects-implant-break}
\end{figure}

The fact that the widths of the implant and aluminum strips differ it is possible to inspect the implant strip underneath the Al strip due to the layer topology. Here, the amount of light reflected into the camera shows the edges of this complex 3D structure.
Figure~\ref{fig:defects-implant-break} shows a detection example of the implant break defect type. The left-hand side panel shows the source image. The right-hand side panel shows the implant strip break detected. The detection is done by recognizing the strips with pattern matching and using the edge detection algorithms over the detected strips. The fact that edges are detected indicates a defect along the green edge detection line.

\begin{figure}[!htb]
	\hspace*{\fill}
	\includegraphics[width=.70\textwidth]{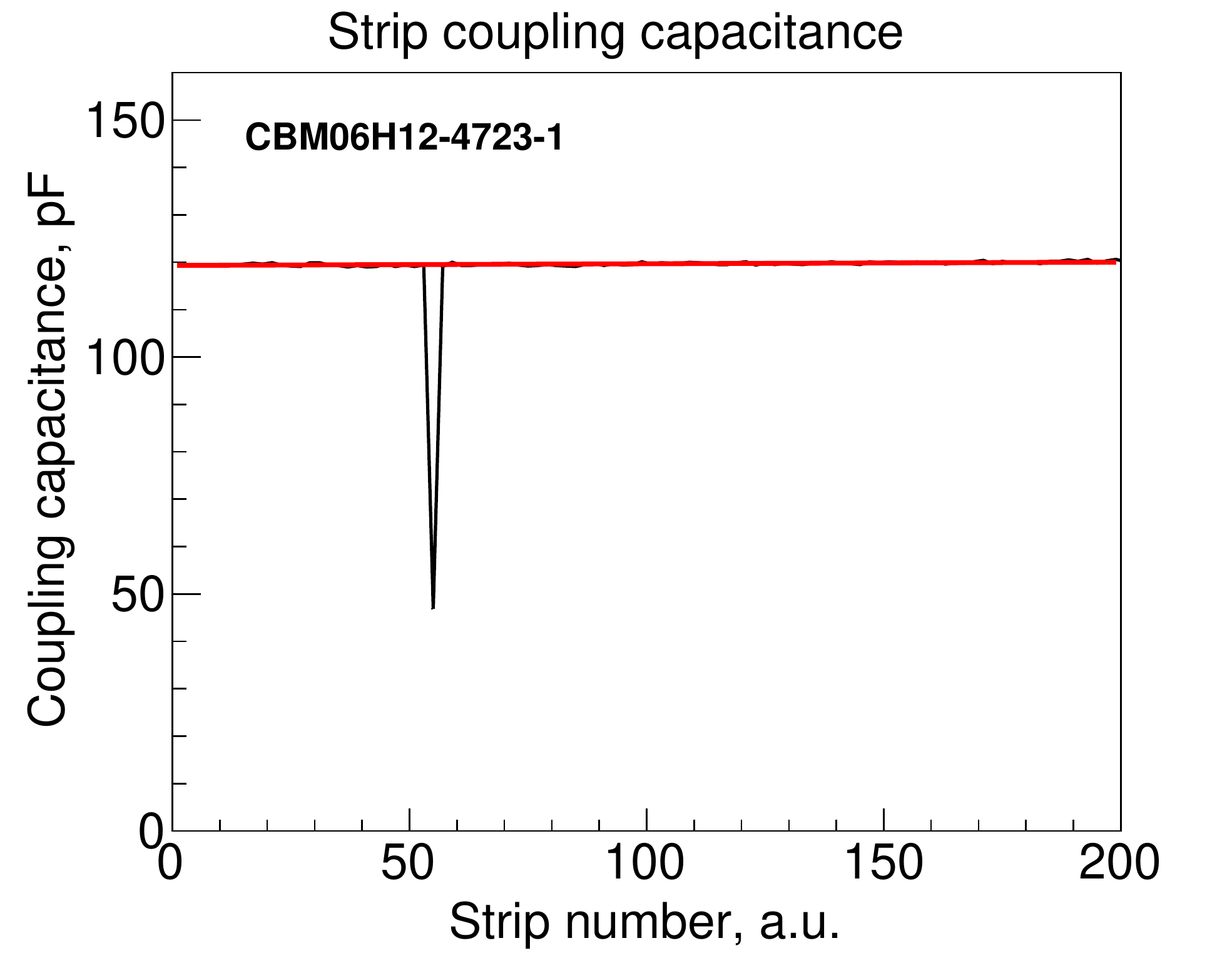}
	\hspace*{\fill}

	\caption{Strip coupling capacitance measured during the electrical inspection. Low value measured for the strip 58 indicates the defect.}
	\label{fig:defects-implant-break-cc}
\end{figure}

The sensor with implant breaks detected was inspected electrically as well. The measurement sensitive to the implant break defect is the strip coupling capacitance. Fig. \ref{fig:defects-implant-break-cc} shows the strip coupling capacitance measured for the prototype sensor. The lower capacitance of the strip 58, compared to the neighbors, verifies the defect observed during optical inspection.

Further correspondence between various defects found optically and measured electrical properties of the sensors is demonstrated in \cite{Lavrik2017}. Thus, the optical inspection system has high predictive power and is able to identify the sensors for further thorough electrical measurements. A more detailed account on the correlation of optical defects and  electrical sensor properties will be subject of a forthcoming publication based on Refs.~\cite{Panasenko2016, Panasenko2018}.

The above examples show the power of the employed algorithms. With proper tuning, they can be applied to various classes of defects. In our case, i.e., for the sensors foreseen for the CBM STS detector, the algorithms were adjusted to find specific defects such as implant shorts, p-stop implant defects, dielectric defects, passivation layer defects, as well as edge defects. The latter ones are defects where small pieces of silicon are chipped from the edges during the cutting process.
The investigated sensor's regions comprise not only the strips but also more \textgravedbl complicated\textgravedbl\ regions, i.e., guard and bias rings, bonding pads and bias resistors.

Overall, 25 sensor samples were characterized. They were picked from all major sensor generations, sizes and metalization types: double metalization and single metalization (phased out historical version). For this limited number of sensors, optical inspection was performed manually in addition. This allowed to compare the total amount of the defects present and the number of defects detected with automated algorithms.

For the samples investigated with the methods described above, we were able to find and categorize correctly on average 87\% of the defects, i.e. their type and severity levels were assigned correctly.

\section{Advanced Image Analyses}
\label{Advanced_Image_Analysis}

The methods described above provide a powerful set of defect detection algorithms. However, they are configuration dependent, i.e. provide insufficient result if not properly adjusted. For advanced defect analysis, we investigate the possibility of neural network applications for the defect detection and classification.

We use convolutional deep neural networks~\cite{Lecun1998} for defect detection and classification. We implement defect detection and  context detection networks separately to reduce the misidentification, since the object to be detected can be overlaid. The networks employed here are based on the Darknet~\cite{Redmon2016} framework implementing the modified Faster R-CNN~\cite{Girshick2016} network model. It allows to detect objects which look similar to those trained against, as well as to predict the object bounding box, e.g. its location on the source image.

\begin{figure}[!htb]
	\centering
	\includegraphics[width=.99\textwidth]{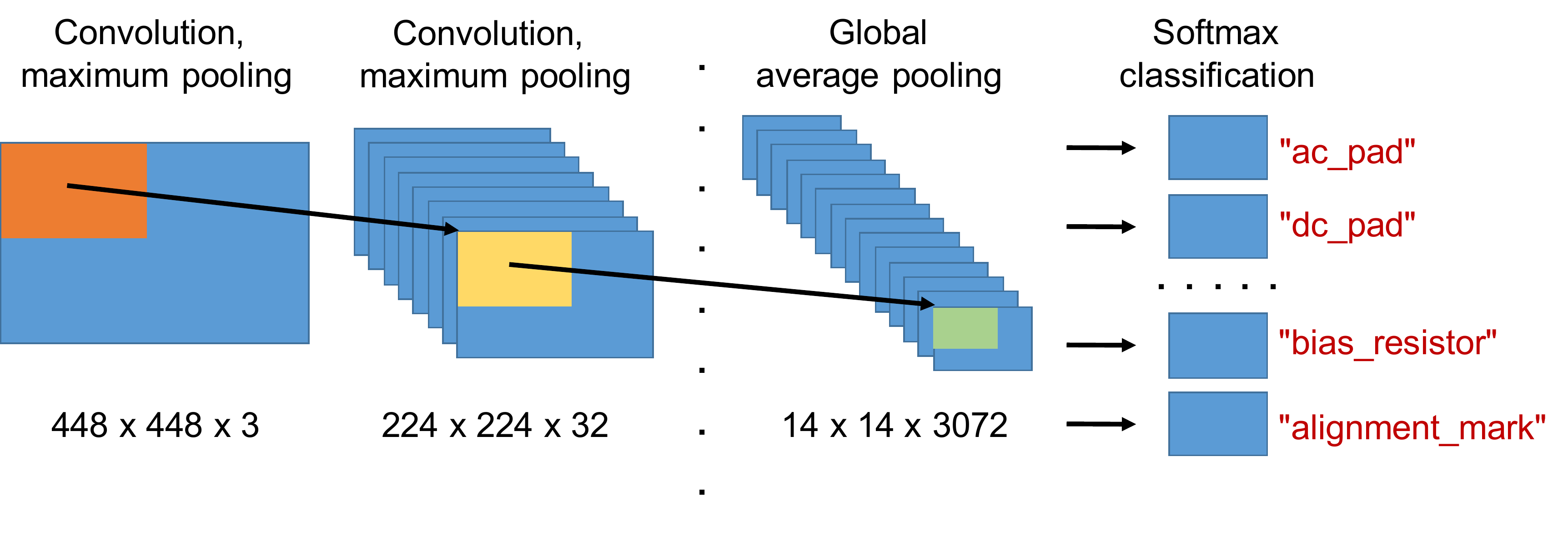}
	\caption{A simplified sketch of the neural network architectures used in this work.}
	\label{fig:neuro-network}
\end{figure}

Figure~\ref{fig:neuro-network} shows a simplified sketch of the deep convolutional neural networks used in this work, the full network architecture is provided in~\cite{Lavrik2017}. It features many convolutional layers, which apply image convolution operations with certain amount of filters with different convolution kernels. The convoluted images retain their size but the depth of the feature maps is modified according to the amount of filters applied. Each convolution layer is activated with an activation function which governs the propagation of the information to the next layers of the network. Examples of these functions are sigmoid, hyperbolic tangent, rectified linear unit and parametrized rectified linear unit functions~\cite{He2015PReLU}. In this work, we use parametrized rectified linear unit functions which, for deep neural networks, provides better classification accuracy compared to the others~\cite{He2015PReLU}.

The maximum pooling layers extract the maximum pixel values of the filtered images within a certain evaluation window and propagate them to the next layers of the network, thus decreasing the size of the images.

By repeating convolutions and maximum pooling, high-dimensioned feature maps are obtained (in this work up to $14 \times 14 \times 3072 \approx 6 \cdot 10^5$ features).

The extracted feature maps are analyzed for the probability of a certain class of object to appear based on the softmax logistic function~\cite{Bishop2006}:

\begin{equation}
\label{eq:softmax}
p(C_k|x) = \frac{exp(a_k)}{ \Sigma_j exp(a_j)},
\end{equation}

which allows to compute the probability $p(C_k|x)$ of certain object class $k$ to be observed among $j$ total classes defined.

The output of the optical inspection was taken as source data for the neural networks training: for the context-detection source data, the pattern matching algorithms with lowered detection threshold values were used; for the defects, the results of the morphological analysis were used. In both cases, the location and size of the feature were used to automatically form the input for the neural network.

The training was done on an OEM Nvidia GTX 745 GPU with 2000 MB of RAM available and CUDA~\cite{CUDA} support. The full 5~MP source images can not be used due to the memory and performance limitations of the GPU. Instead, the regions of interest around a defect or feature of about $450 \times 450$ pixels were extracted for further processing. The models were trained for 10000 iterations taking about 5 days for each model. The data sets were automatically augmented by the framework by random source image rotation, crop and resizing. This allows to improve the robustness of the detection by increasing the generalization of the feature detection.

The Faster-RCNN~(\cite{Girshick2015}, \cite{Girshick2016}) neural network model used in this work allows both to classify the object detected and to propose the region in the image where the object is located. To qualify the performance of the network, the sum of precision of both metrics is taken into account. Faster-RCNN defines the optimization function, or the loss functions as:

\begin{equation}
\label{eq:rcnn-loss}
L(\{p_i\}, \{t_i\}) = \frac{1}{N_{cls}} \sum_i L_{cls} (p_i^{}, p_i^*) + \lambda \frac{1}{N_{reg}} \sum_i p_i^* L_{reg} (t_i^{}, t_i^*),
\end{equation}

where $L_{cls}$ and $L_{reg}$ are loss terms related to the object class and object region identification~\cite{Girshick2016}.

To measure the quality of the bounding regions of the detected objects proposed by the Faster-RCNN, various metrics can be used such as precision, recall and intersection over union (IOU). The \textit{precision} metric is the ratio between the intersection area of the predefined source object region (so called ground truth) and the detected object region area. The \textit{recall} metric defines the ratio of the intersection region and the source object. The \textit{IOU} metric denotes the ratio between the region of the intersection area of the source object and the detected object and the union of their areas. We use the latter parameter to estimate the performance of the model in this work.

\begin{figure}[!htb]
	\hspace*{\fill}
	\includegraphics[width=.49\textwidth]{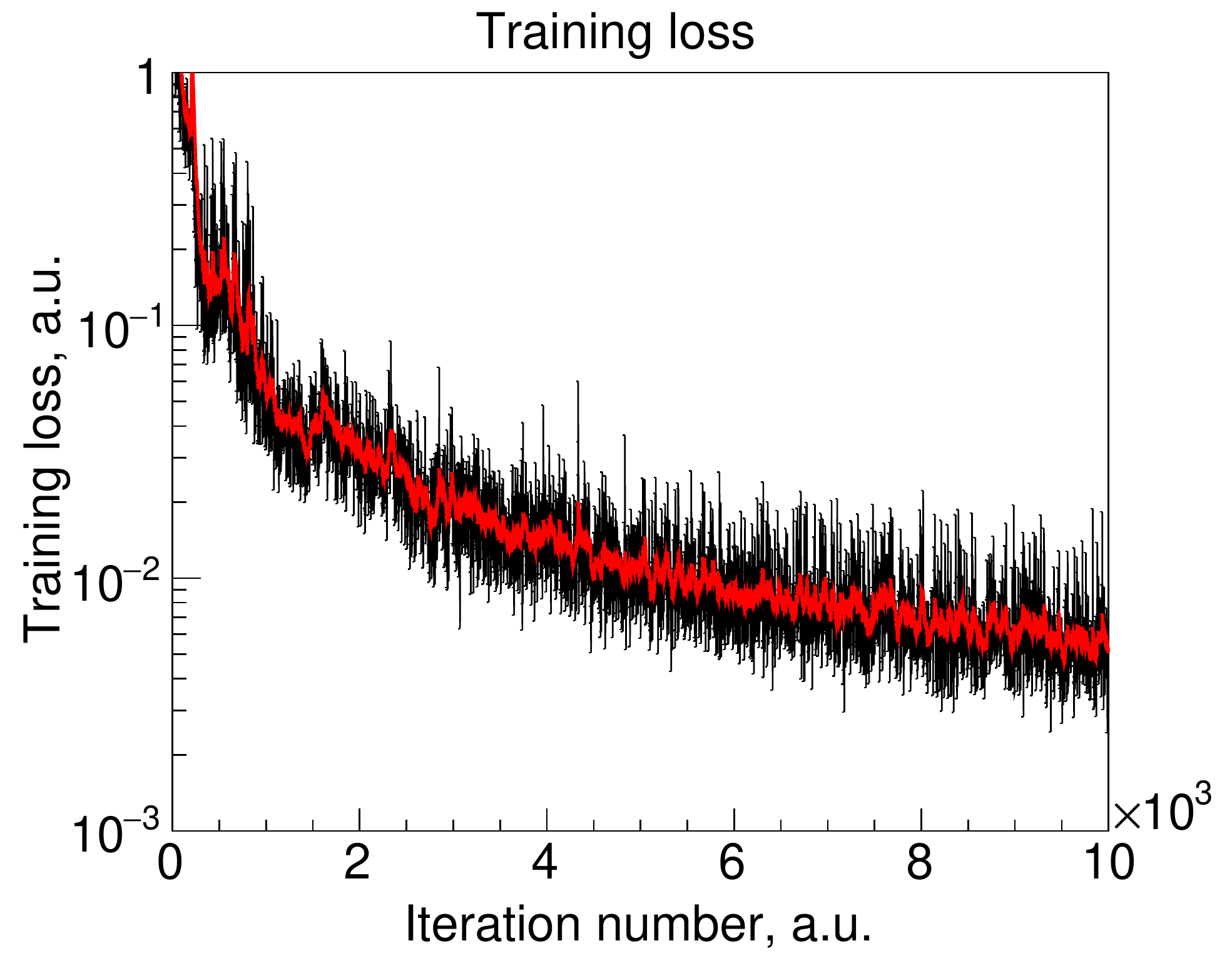}
	\hfill
	\includegraphics[width=.49\textwidth]{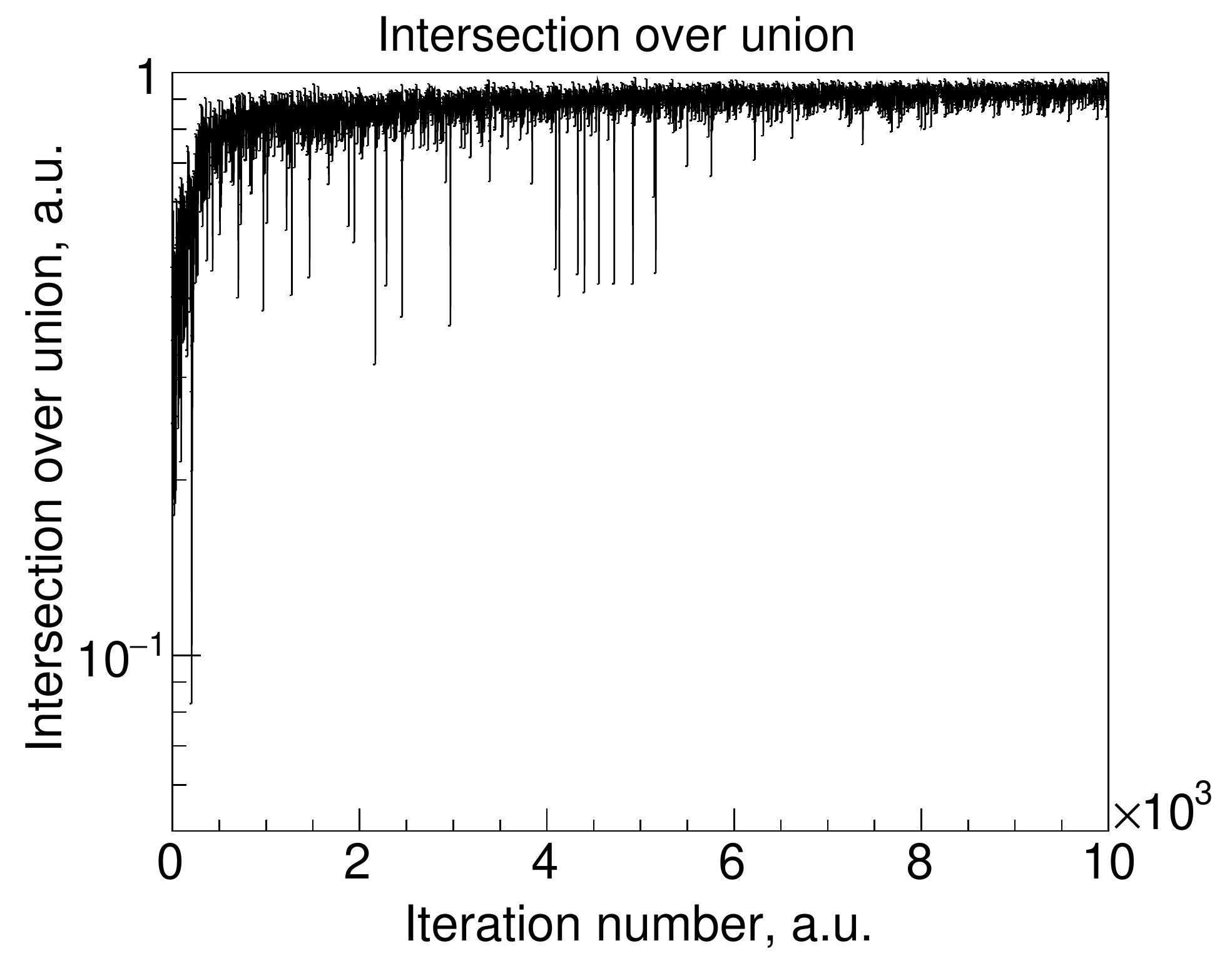}
	\hspace*{\fill}
	\caption{Training progress illustrated in terms of training loss (left panel) and intersection over union (right panel).}
	\label{fig:neuro-training}
\end{figure}

Figure~\ref{fig:neuro-training} illustrates the progress of the training of the neural network for defects detection. The iterations represent a full training data set processing step. The training loss (left panel, black curve) represents the sum of errors made for each prediction and is the minimization parameter, the evolution of the average training loss is depicted with a red curve. The \textit{IOU} metric evolution is depicted on the right panel. In a well trained network, this metric should tend to be equal to unity.

\begin{figure}[!htb]
	\hspace*{\fill}
	\includegraphics[width=.30\textwidth]{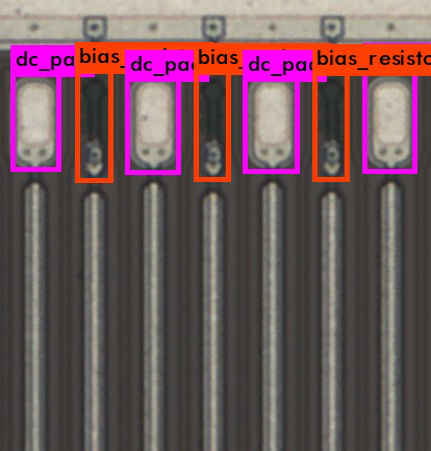}
	\hfill
	\includegraphics[width=.30\textwidth]{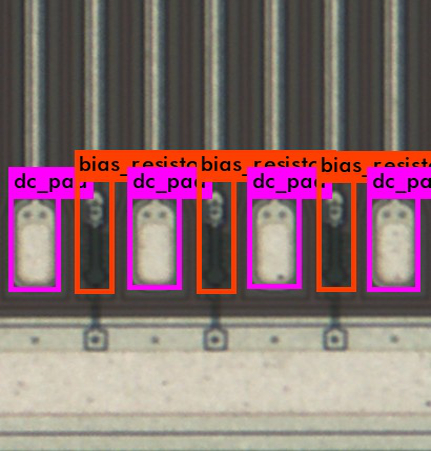}
	\hspace*{\fill}
	
	\rule[-0.3ex]{0pt}{1ex}
	
	\hspace*{\fill}
	\includegraphics[width=.30\textwidth]{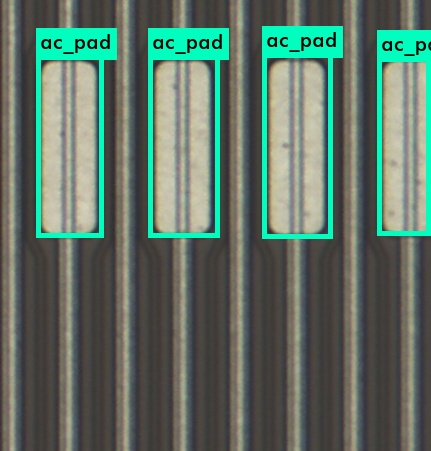}
	\hfill
	\includegraphics[width=.30\textwidth]{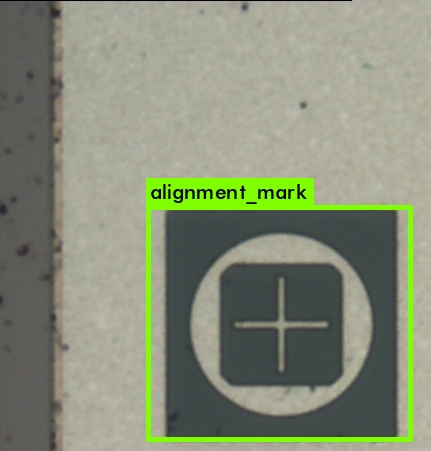}
	\hspace*{\fill}
	\caption{Different contexts of the sensor identified with a neural network.}
	\label{fig:neuro-context}
\end{figure}

Figure~\ref{fig:neuro-context} shows the output of the trained context detection neural network. It allows to identify the electrical elements and other features of the sensor.

\begin{figure}[!htb]
	\hspace*{\fill}
	\includegraphics[width=.20\textwidth]{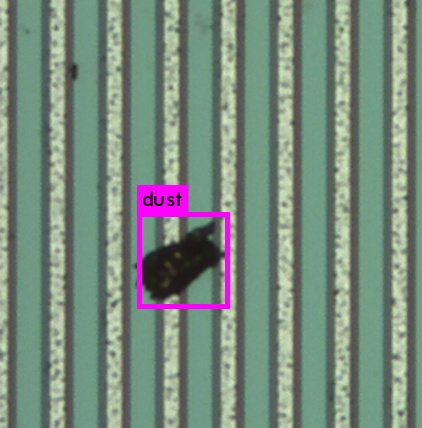}
	\hfill
	\includegraphics[width=.20\textwidth]{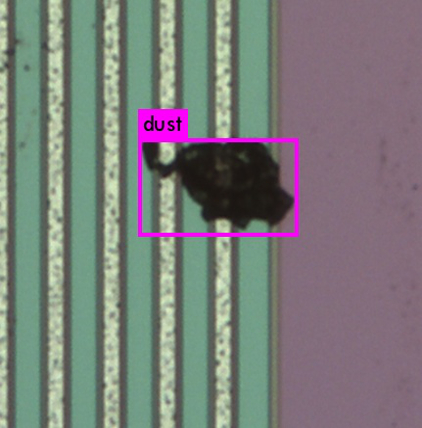}
	\hfill
	\includegraphics[width=.20\textwidth]{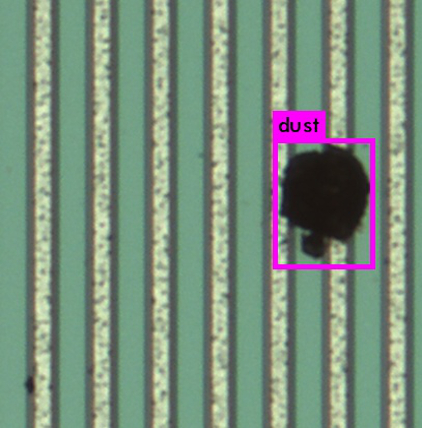}
	\hfill
	\includegraphics[width=.20\textwidth]{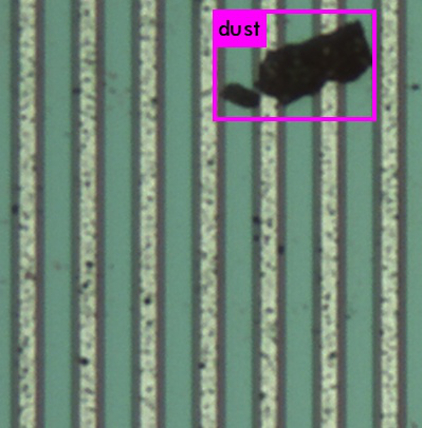}
	\hspace*{\fill}
	
	\rule[-0.3ex]{0pt}{1ex}

	\hspace*{\fill}
	\includegraphics[width=.20\textwidth]{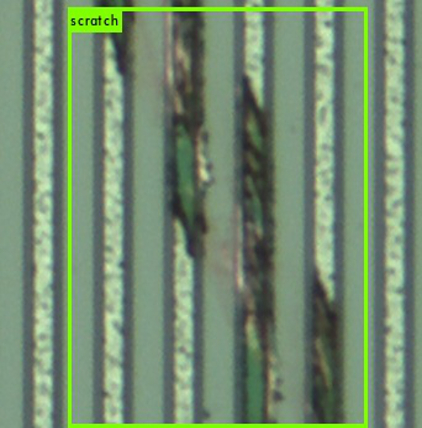}
	\hfill
	\includegraphics[width=.20\textwidth]{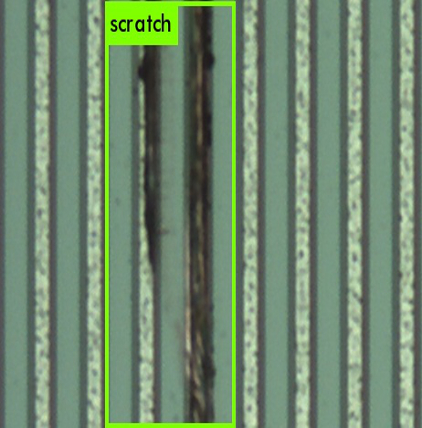}
	\hfill
	\includegraphics[width=.20\textwidth]{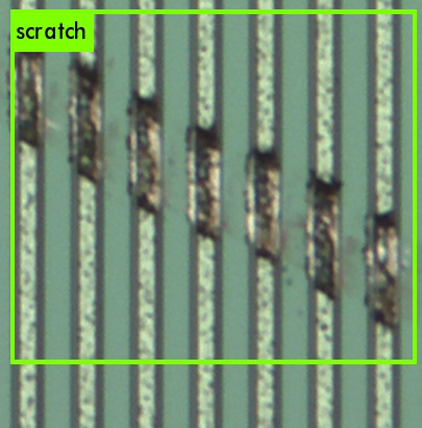}
	\hfill
	\includegraphics[width=.20\textwidth]{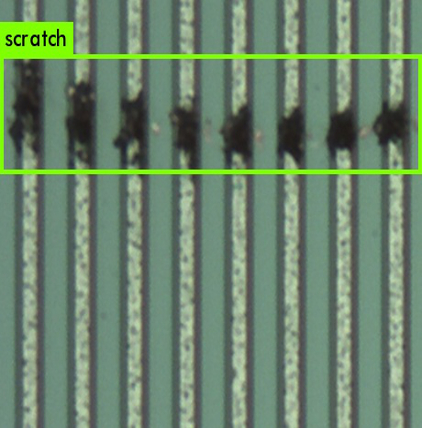}
	\hspace*{\fill}
	\caption{The dust particles and scratches identified with a neural network.}
	\label{fig:neuro-dust}
\end{figure}

The output of the trained defect detection neural network is depicted in Fig.~\ref{fig:neuro-dust}. Here the dust particles and scratches are detected. The associated regions where the defects are found are extracted as well.

The defect detection neural network is continuously trained on the new data from the sensors as they are inspected. This allows to account for the defects which have different spatial configuration.

The GPU used in this work is somewhat outdated compared to the state of the art solutions and has a limiting impact on the processing speed and image size. The network, once trained on the GPU, can be run for the defect finding and classification on the CPU as well. A single $448 \times 448$ source image takes about 0.16 seconds to be processed on the GPU and about 7.9 seconds to be processed on the CPU. A single 5 megapixel image taken from the camera results in 30 source images for the network input, yielding a total time of about 5 seconds for the GPU and about 4 minutes for the CPU based detection.

The usage of the neural networks has improved the defect detection to 96\% and the defect context detection to 93\%.

\section{Summary}
We have developed a setup to automatically scan large sets of silicon microstrip sensors for defects like scratches, broken readout strips and many others. The sensors are foreseen for the large area CBM silicon tracking system. Microscopic images of the sensors are analyzed employing Machine Vision Algorithms from NI Vision package with high precision. The software is able to recognize and classify defects with a detection rate $>$~90\%. We have demonstrated that advanced image processing based on neural network techniques is able to further improve the recognition and defect classification rate. The recognition via Neural Networks is much less dependent on the configuration of the software.

It is important to correlate the found optical defects to  full chain detector tests to be able to classify the severity of optically found defects. At present, the final readout ASICs are not yet available. As soon as the full readout will become available these test are scheduled.

We should finally remark that the techniques developed here are not restricted to the silicons sensors. The system can be adapted to almost any problem, where regular, well defined microscopic structures need to be scanned for defects.

\section*{Acknowledgements}
The work was performed under grant BMBF 05P16VTFC1. We thank our colleagues Dr. J. Heuser and Dr. A. Lymanets, GSI Darmstadt, for fruitful discussions of the subject and for their support when installing the setup.

\section*{References}

\bibliography{literature/bibliography}

\end{document}